\documentclass[copyright,creativecommons]{eptcs}
\providecommand{\event}{DCM - FLoC  2010} 

\usepackage{breakurl}             
\usepackage{color}
\usepackage{graphicx,epsfig}

\newcommand{\tr}{\mbox{tr}}
\renewcommand{\d}{\mbox{d}}

\title{Landauer's principle in the quantum domain}
\author{Janet Anders \qquad\qquad Saroosh Shabbir
\institute{Dept of Physics \& Astronomy}
\institute{University College London\\
London WC1E 6BT, UK}
\email{j.anders@ucl.ac.uk}
\and
Stefanie Hilt \qquad\qquad 	Eric Lutz
\institute{Department of Physics}
\institute{University of Augsburg \\
86135 Augsburg, Germany}
}
\def\titlerunning{Landauer's principle in the quantum domain}
\def\authorrunning{J. Anders, S. Shabbir, S. Hilt \& E. Lutz}
\begin{document}
\maketitle

\begin{abstract}

Recent papers discussing thermodynamic processes in strongly coupled quantum systems claim a violation of Landauer's principle and imply a violation of the second law of thermodynamics \cite{all00,all01,leff03, hoe05}. If true, this would have powerful consequences. Perpetuum mobiles could be build as long as the operating temperature is brought close to zero. It would also have serious consequences on thermodynamic derivations of information theoretic results, such as the Holevo bound \cite{plen99}.  Here we argue why these claims are erroneous. Correlations occurring in the strongly coupled, quantum domain require a rethink of how entropy, heat and work are calculated. It is shown that a consistent treatment solves the paradox \cite{HALS}.

\end{abstract}

\section{Landauer's principle}

Computers are heat engines that take energy, $E$, and ``work out'', $W$,  the solution to a problem. Real computers however also dissipate heat, $Q$. This wasted energy is exactly the difference between what the computer is fed and it's productive output work, $E - W = Q$. This fundamental balance equation is known as the first law of thermodynamics. In information theory, in the 60ties, workers started to wonder if the computer's efficiency could be improved by dumping an arbitrarily small amount of heat, i.e. could a computer  run without dissipating any heat? The answer is {\it no}! While many computational operations can be run (reversibly) without generating heat, there is a fundamental limit to the heat released when \emph{one bit of information is erased}. Landauer's principle  \cite{lan61,lan88} sets the lower bound to this heat as $k_B T \ln 2$ when the erasure takes place at temperature $T$, and with $k_B$ the Boltzmann constant. The erasure principle was orginally formulated for classical bits. Lubkin \cite{lub87} generalised the erasure to general quantum states $\rho$\footnote{$\rho$ are hermitian, positive operators with $\tr[\rho] =1$.}  erasing which will generate a heat of at least
\begin{equation}
	Q_{erasure} \ge k_B T \, S(\rho),
\end{equation}
where $S(\rho) = - \tr[\rho \ln \rho]$ is the von Neumann entropy of $\rho$.

\medskip
Landauer's principle plays an important role in the resolution of a 19th century thought experiment, involving a box filled with gas and a demon \cite{leff90,Maxwelldemon}. The demon, known as Maxwell's demon, can follow the motion of the gas particles and separate faster particles from slower ones, thus effecting two reservoirs with different temperatures. He can then extract work from the gas by placing a piston in the way of fast gas particles. The fast particles would push the piston (= work) which in turn slows them down (= reduce heat), with no change of total energy, i.e. $W = - Q$. This process would run contrary to the second law of thermodynamics which states that no cyclic process can convert heat completely into work. The paradox can be resolved by realising that for the thermodynamic cycle to be closed, any \emph{information} on the gas particles stored by the demon has to be erased! By  Landauer's principle this \emph{generates} heat, and a careful analysis shows that it is just the right amount to tame the demon's paradox \cite{lan61,leff90,benn82}.  

\section{Information theoretic bounds from Landauer's principle}

Landauer's principle can be used to prove important theorems in quantum information theory, for instance, the Holevo bound on the accessible information and the no cloning theorem \cite{plen99,plen01,koji09}. For the derivation of the Holevo bound consider a sender, Martin, that sends a sequence of quantum states $\rho_i$, randomly chosen with probability $p_i$. The receiver, Amy, is told what Martin's states are and their likelyhood of appearing, however, she does not know which of these states she is receiving. Her ignorance about the actual choice of each copy sent means that the state she writes down on her side is $\rho = \sum_i p_i \rho_i$. 

Now Amy wants to find out what sequence of states $\rho_i$ Martin sent. She can make any measurement she likes on her state $\rho$ to find as much information as she can about Martin's choices. The number of bits that Amy can decode by a measurement $M$ on $\rho$ is the \emph{mutual information}, 
\begin{equation}
	I_M =   \sum_{i, m} \, p_{im}	\, \ln {p_{im} \over p_{i} \, q_{m}},
\end{equation}
where $p_{im}$ are the joint probabilities of Martin sending $\rho_i$ and Amy finding the $m$-th measurement outcome, while  $q_m = \sum_i p_{im}$ and $p_i = \sum_m p_{im}$ are the marginal probabilities of Amy finding outcome $m$ and Martin sending $\rho_i$. Amy's choice of measurements will clearly depend on the given preparation procedure, $\rho_i$ with $p_i$. By optimising her measurements accordingly the maximum information she can gain is 
\begin{equation}
	I_{acc} = \max_M \, I_M.
\end{equation}
This number of bits is the \emph{accessible information} and quantifies how much classical information Martin and Amy share.  

To find an upper bound on the accessible information we will use Landauer's principle \cite{plen99,plen01,koji09}. According to Landauer's principle, erasing the information on Martin's side generates heat. Martin had initially the set of states $\rho_i$, each containing $S(\rho_i)$ bits of classical information, erasure of which would generate at least $k_B T S(\rho_i)$ per state. Erasing Martin's ensemble will thus release the minimal  heat of  
\begin{equation}
	Q_{erase}^{Martin} = \sum_i \, p_i \, \, k_B T S(\rho_i),
\end{equation}
per copy sent. On the other side, Amy receives a state with information content $S(\rho)$ erasing which would require a heat generation of at least 
\begin{equation}
	Q_{erase}^{Amy} = k_B T S(\rho). 
\end{equation}
Now Amy's minimal erasure heat is higher than Martin's because she is more ignorant about what state she has, $S(\rho) \ge \sum_i \, p_i \, S(\rho_i)$. Energy conservation requires that the difference in erasure must be somewhere! Indeed it can only be associated with the \emph{shared information}, at best $I_{acc}$, between Martin and Amy. This can be stated as 
\begin{equation}
 	Q_{erase}^{Amy} - Q_{erase}^{Martin}  = Q_{erase}^{shared},
\end{equation}
where $Q_{erase}^{shared} \ge k_B T \, I_{acc}$ is the heat that would be released if the shared information was erased. Rearranging leads to
\begin{equation}
	 I_{acc} \le  {Q_{erase}^{shared} \over k_B T} 
	 	= S(\rho) -  \sum_i \, p_i \, \, S(\rho_i),
\end{equation}
which is the Holevo bound on the accessible information\footnote{Note, that if Martin generated heat in excess of the minimal heat of $Q_{erase}^{Martin}$ this would only reduce the shared information. However, the contrary is not true - Amy cannot increase the shared information by deliberately dumping more than the minimum heat of $Q_{erase}^{Amy}$!}. 

\section{Second law of thermodynamics, Clausius inequality and Landauer's principle}

Imagine a ship that powers its engines solely from converting heat contained in the water of the sea it is moving in. To balance the energy bill the sea would get a bit colder but this would be almost compensated again by the friction between the ship's moving body and the water. What a fabulous way to travel! This marvellous engine would solve all our energy problems simply by being able to convert heat to ``productive energy''. Unfortunately this machine, known as \emph{perpetuum mobile of the second kind}, is forbidden by a fundamental law of the physics, the second law of thermodynamics.

\medskip

The laws of thermodynamics describe the transport of heat and work in thermodynamic processes. The first law states that energy can neither be created nor destroyed, it can only change forms. Therefore in an isolated system, the total energy remains the same in any process. Whereas the first law of thermodynamics describes simply the conservation of energy for a thermodynamic system, the second law makes a prediction about how the different forms of energy can be transformed into each other. There are many formulations of the second law of thermodynamics, but all are equivalent in the sense that each form of the second law logically implies every other form. A descriptive one was already given above. It is called the heat engine formulation by Lord Kelvin: It is impossible to convert heat completely into work in a cyclic process. With the introduction of the entropy by Clausius a mathematical description of the second law, the so-called Clausius inequality, was possible. It asserts that, for a system initially in a thermal state, the heat $Q$ received  by the system at temperature $T$ cannot exceed the system's change of entropy, $\Delta S$, multiplied by the temperature, 
\begin{equation}
    Q \le k_B T \Delta S,
\end{equation}
where the equality sign holds for quasistatic, reversible processes.

\medskip

From Clausius' inequality it is easy to derive Landauer's principle \cite{HALS}. Consider an isolated system with two stable states that are used to encode one bit of information (for instance a symmetric double-well potential with high energy barrier). The system is initially in equilibrium at temperature $T$ and the two states are occupied with equal probability. We reset the memory by first coupling it to the reservoir and then modulating the potential in order to bring the system with probability one into one of its states \cite{lan61}. The von Neumann entropy of the system is hence $\ln 2$ before the coupling to the reservoir and zero after complete erasure. From the Clausius inequality, we then find that the dissipated heat obeys 
\begin{equation}
	Q_{dissipated} = - Q \ge - k_B T \Delta S = k_B T \ln 2. 
\end{equation}
This is Landauer's erasure principle.

\section{Violation of Clausius' inequality} 

Recently, there has been a revived interest in Landauer's principle when applied to quantum systems. A number of publications \cite{all00,all01,hoe05,leff03} have argued, using a model of a quantum Brownian oscillator interacting with a bath, that the second law of thermodynamics can be violated. For a thermodynamic process in which the oscillator's mass is increased it was shown that the Clausius inequality is not fulfilled, i.e. $Q \not \le k_B T \Delta S$, implying also that it would be possible to violate Landauer's principle. 

Here is a short summary of the arguments leading to the violation. The starting point of investigation is the Hamiltonian of the oscillator with mass $M$ and oscillation frequency $\omega$, 
\begin{equation}
	H_{o} = {p^2 \over 2M} + {M \omega^2 q^2 \over 2}.
\end{equation} 
$p$ and $q$ are the momentum and position operators of the oscillator and they fulfil the commutation relations $[q, p]= i \hbar$. When the oscillator is coupled to a bath the Hamiltonian will adopt the form 
\begin{equation}
	H = H_o + H_b + H_{ob}
\end{equation}
where the $H_b$ is the bath Hamiltonian and $H_{ob} (\eta)$ describes the coupling parameterised by $\eta$ between oscillator and bath\footnote{Details of the Hamiltonian and further calculation can be found in standard text books \cite{wei99}  as well as in the paper discussing the violation \cite{all00} and the paper resolving the paradox \cite{HALS}.}. The equilibrium state of oscillator and bath at temperature $T$ is given by
\begin{equation}
	\rho_{ob} = {e^{- {H \over k_B T}} \over \tr[e^{- {H \over k_B T}} ]},
\end{equation}
the quantum mechanical analogue of the Boltzmann distribution. Expectation values of the oscillator operators, $O_o$, are evaluated as $\langle O_o \rangle = \tr[O_o \, \rho_{ob}] = \tr_o[O_o \, \rho_{o}]$ where $\rho_o = \tr_{b}[\rho_{ob}]$ is the reduced state of the oscillator. The characteristics of the oscillator are its mean position  and momentum 
\begin{equation}
	\langle q\rangle =0 \quad \mbox{and} \quad \langle p \rangle =0,
\end{equation}
and their variations, 
\begin{equation}
	\langle q^2 \rangle - \langle q \rangle^2 = : f_1(T, \omega, M, \eta) 
		\quad \mbox{and} \quad 
	\langle p^2 \rangle - \langle p \rangle^2  = : f_2(T, \omega, M, \eta),
\end{equation}
which are known functions of the temperature, the frequency and mass of the oscillator and the coupling strength between oscillator and bath (for details see \cite{HALS}).

The entropy of the oscillator can readily be expressed as a function of these moments, 
\begin{equation}
	S (T, \omega, M, \eta) = \left(v + {1 \over 2}\right) \ln \left(v + {1 \over 2}\right) - \left(v - {1 \over 2}\right) \ln \left(v - {1 \over 2}\right),
\end{equation}
where $v = v(T, \omega, M, \eta) = \sqrt{f_1(T, \omega, M, \eta) f_2(T, \omega, M, \eta)}/ \hbar$. Changing a thermodynamic parameter $\alpha$ from $\alpha_0$ to $\alpha_1$ implies the entropy change 
\begin{equation}  \label{eq:entropychange}
	\Delta S^{(\alpha)}  = \int_{\alpha_0}^{\alpha_1} \d \alpha  \, {\d S \over \d \alpha}.
\end{equation}
The heat that results from a change of the functions $f_1$ and $f_2$ with the parameter $\alpha$ is given by
\begin{equation} \label{eq:heatchange}
	Q^{(\alpha)} = \int_{\alpha_0}^{\alpha_1} \d \alpha \left[ {1 \over 2 M} {\d f_2 \over \d \alpha} +{ M \omega^2 \over 2}  {\d f_1 \over \d \alpha}\right].
\end{equation}
Finally, when the thermodynamical process is considered in which the oscillator acquires mass, $M_0 \to M_1 > M_0$, the entropy change and heat can be calculated straightforwardly \cite{all01, HALS} with Eqs. (\ref{eq:entropychange}) and (\ref{eq:heatchange}). For low temperatures and strong coupling one finds, for this particular process,
\begin{equation} \label{eq:heatentropyM}
  	\Delta S^{(M)} < 0 \quad \mbox{and} \quad Q^{(M)}  > 0,
\end{equation}
implying 
\begin{equation}
	Q^{(M)} \not \le   k_B T\Delta S^{(M)},
\end{equation}
clearly contradicting Clausius' inequality \cite{all00,all01}. 

\section{Resolution of the paradox} 

This result is, given the connection between Clausius' inequality, the second law of thermodynamics and Landauer's principle, most discomforting! It would have dramatic, science-fiction like consequences for both, thermodynamics and quantum information theory, as it opens the possibility for the existence of heat engines that run just by converting heat into work and information sharing beyond the Holevo, no-cloning and no-signalling. The only requirement to harvest this unbelievable power would be to bring a quantum system into a strong coupling regime and to cool it to low temperatures. Then all these paradoxical processes could be run! So, something must be wrong \cite{for06}! How can we resolve this conundrum?

Rechecking the mathematical steps leading to Eqs.~(\ref{eq:heatentropyM}) one can verify their correctness. However, the reasoning leading to the claim of the Clausius violation is flawed. The culprit in this argument is that the coupling between the oscillator and the equilibrating bath is \emph{not negligible}, as commonly assumed. Indeed, the Clausius inequality assumes that the starting state of the parameter changing process under consideration is a thermal state. When the coupling between oscillator and bath is negligible, i.e. $\eta \to 0$ the reduced state of the oscillator, $\rho_o$ is indeed thermal, i.e. it can be approximated as
\begin{equation}
\label{gibbs}
	\rho_o (T) \approx {e^{- {H_o \over k_B T}} \over \tr[e^{- {H_o \over k_B T}}]}.
\end{equation}
These thermal states are characterised by state functions  $f_1$ and $f_2$ that fulfil (approximately) the relations 
\begin{equation}
	f_1 \, f_2 		\approx \left({\hbar \over 2} \coth {\hbar \omega \over 2 k_B T} \right)^2 
		\quad \mbox{and} \quad 
	{f_2 \over f_1} 	\approx \left(M \omega \right)^2.
\end{equation}
However, when the coupling is strong, the reduced state, $\rho_o$, {\emph deviates} from  the thermal Gibbs form \ref{gibbs}. The state is ``squeezed'' in the sense that the variance in position may be much smaller than usual at the expense of the momentum distribution.

\medskip

A consistent treatment of this situation must include the coupling between the oscillator and the bath \cite{HALS}. The process of increasing the oscillator mass, considered above,  is then evidently a two step process. To compute the entropy change and heat, we first consider the oscillator alone, in a thermal state, $\rho_o (T)$, at temperature $T$. Then it is coupled to the bath quantified by the coupling parameter $\eta$ where now the entire system, oscillator and bath, is again in a thermal Gibbs state, $\rho_{ob}$, at the same temperature $T$, while the reduced state of the oscillator is a non-thermal, squeezed state. Only then is the mass increased as the finally step. The entropy change and heat for this thermodynamically complete process are then 
\begin{equation}
	\Delta S = \Delta S^{(\eta)} + \Delta S^{(M)}  \ge 0 
			\quad \mbox{and} \quad 
	Q = Q^{(\eta)} + Q^{(M)}  \le 0.
\end{equation}
It can further be shown that for all parameter ranges it is
\begin{equation}
	Q  \le k_B T \Delta S,
\end{equation}
which resolves the paradox and re-establishes the validity of the Clausius inequality, and hence Landauer's principle, in the quantum domain \cite{HALS}.

\section{Discussion} 

The lesson to be learnt from the paradoxical situation is that when correlations are strong (both classical and quantum), arguments and proofs that are based on thermodynamical calculations have to be carefully revised. Standard thermodynamics assumes that the correlations, between the system to be described and the equilibrating bath, scale with the surface area of the system and that this contribution is negliable compared to the volume of the system. This is not true in many situations! Small systems, such as a single oscillator, do clearly not obey this logic. Moreover, quantum correlations in realistic models will indeed show entanglement that scales with the surface area (known as \emph{area laws}). However, the extraordinary strength of these particular correlations may still imply that their overall contribution scales with the volume of the system. In all these cases the correlations have to be treated explicitly and their thermodynamical value needs to be taken into account! Neglecting this will invariably result in erroneous conclusions and paradoxes. 

\medskip

{\bf Acknowledgements:} JA thanks the Royal Society (London) for support in form of a Dorothy Hodgkin Fellowship. This work was further supported by the Emmy Noether Program of the DFG (Contract LU1382/1-1) and the cluster of excellence Nanosystems Initiative Munich (NIM).

\vfill

\bibliographystyle{eptcs} 

\end{document}